\documentclass[amsmath,prb,twocolumn,superscriptaddress]{revtex4}
\usepackage{color}
\usepackage{amsfonts}
\usepackage{amssymb}
\usepackage{graphicx}
\usepackage{pstricks}

\begin{document}
\title{Auxiliary master equation for nonequilibrium dual-fermion approach}
\author{Feng Chen}
\affiliation{Department of Physics, University of California San Diego, La Jolla, CA 92093, USA}
\author{Guy Cohen}
\affiliation{The Raymond and Beverley Sackler Center for Computational Molecular and Materials Science, Tel Aviv University, Tel Aviv 69978, Israel}
\affiliation{School of Chemistry, Tel Aviv University, Tel Aviv 69978, Israel}
\author{Michael Galperin}
\email{migalperin@ucsd.edu}
\affiliation{Department of Chemistry \& Biochemistry, University of California San Diego, La Jolla, CA 92093, USA} 

\begin{abstract}
We introduce auxiliary quantum master equation -  dual fermion approach (QME-DF)
and argue that it presents a convenient way to describe steady-states of correlated
impurity systems. The combined scheme yields an expansion around a reference
much closer to the true nonequilibrium state than in the original dual fermion formulation.
In steady-state situations, the scheme is numerically cheaper and allows to avoid long time propagation
of previous considerations. Anderson impurity is used as a test model.
The QME-DF simulations are compared with numerically exact tdDMRG results.
\end{abstract}

\maketitle

Since its theoretical invention~\cite{aviram_molecular_1974} and first experimental evidence of possibility 
to make measurements on single-molecule junctions~\cite{reed_conductance_1997}, 
molecular electronics challenges theory for proper description of response in open molecular systems 
far form equilibrium. 
Often, theoretical treatments are based on perturbative expansion in small parameter
which is usually chosen as strength of intra-molecular interactions or molecule-contacts
couplings. The former can be conveniently treated within the standard nonequilirbium 
Green function (NEGF) technique~\cite{haug_quantum_2008,stefanucci_nonequilibrium_2013}, 
while the latter is described using tools of nonequilibrium atomic limit~\cite{white_nonequilibrium_2014} 
such as,  e.g., many-body flavors of Green function methodology: 
pseudo-particles (PP)~\cite{oh_transport_2011,aoki_nonequilibrium_2014} or 
Hubbard NEGF~\cite{chen_nonequilibrium_2017,miwa_towards_2017} techniques.
These two limits account for majority of experimental measurements.
For example, inelastic electron tunneling spectroscopy~\cite{wang_inelastic_2004}
is usually treated within NEGF~\cite{frederiksen_inelastic_2007,avriller_inelastic_2012}, 
while Coulomb blockade~\cite{poot_temperature_2006},
single molecule strong coupling in plasmonic nanocavities~\cite{chikkaraddy_single-molecule_2016}
or coherent electron-nuclear dynamics~\cite{repp_coherent_2010} require many-body states
based analysis~\cite{seldenthuis_vibrational_2008,white_collective_2012}.   

In the absence of small parameter or when molecule-contacts correlations
cannot be adequately described within perturbation theory, theoretical treatment is more involved.
For example, this is the situation one encounters in describing measurements revealing Kondo physics
in molecular junctions~\cite{park_coulomb_2002,liang_kondo_2002,yu_kondo_2005,osorio_electronic_2007,parks_mechanical_2010,wagner_switching_2013,rakhmilevitch_vibration-mediated_2015}.
Theoretical methods for strongly correlated systems include dynamical mean field theory 
(DMFT)~\cite{anisimov_electronic_2010,aoki_nonequilibrium_2014}, 
density matrix renormalization group (DMRG) technique~\cite{schollwock_density-matrix_2005,schollwock_density-matrix_2011},
scattering states-numerical renormalization group approach~\cite{anders_steady-state_2008,schmitt_comparison_2010},
flow equation approach~\cite{wegner_flow-equations_1994,kehrein_flow_2006}, 
multilayer multiconfiguration time-dependent Hartree (ML-MCTDH)
method~\cite{wang_numerically_2009,wang_multilayer_2018}, 
continuous time quantum Monte Carlo (CT-QMC)~\cite{cohen_taming_2015,antipov_currents_2017,ridley_numerically_2018}
and others. Majority of these methods are numerically demanding and mostly can be applied
only to simple models. 
Relatively less demanding DMFT was extensively used in
simulations of strongly correlated materials (extended systems). Main assumption of the method is 
local character of correlations. 

\begin{figure}[b]
\centering\includegraphics[width=\linewidth]{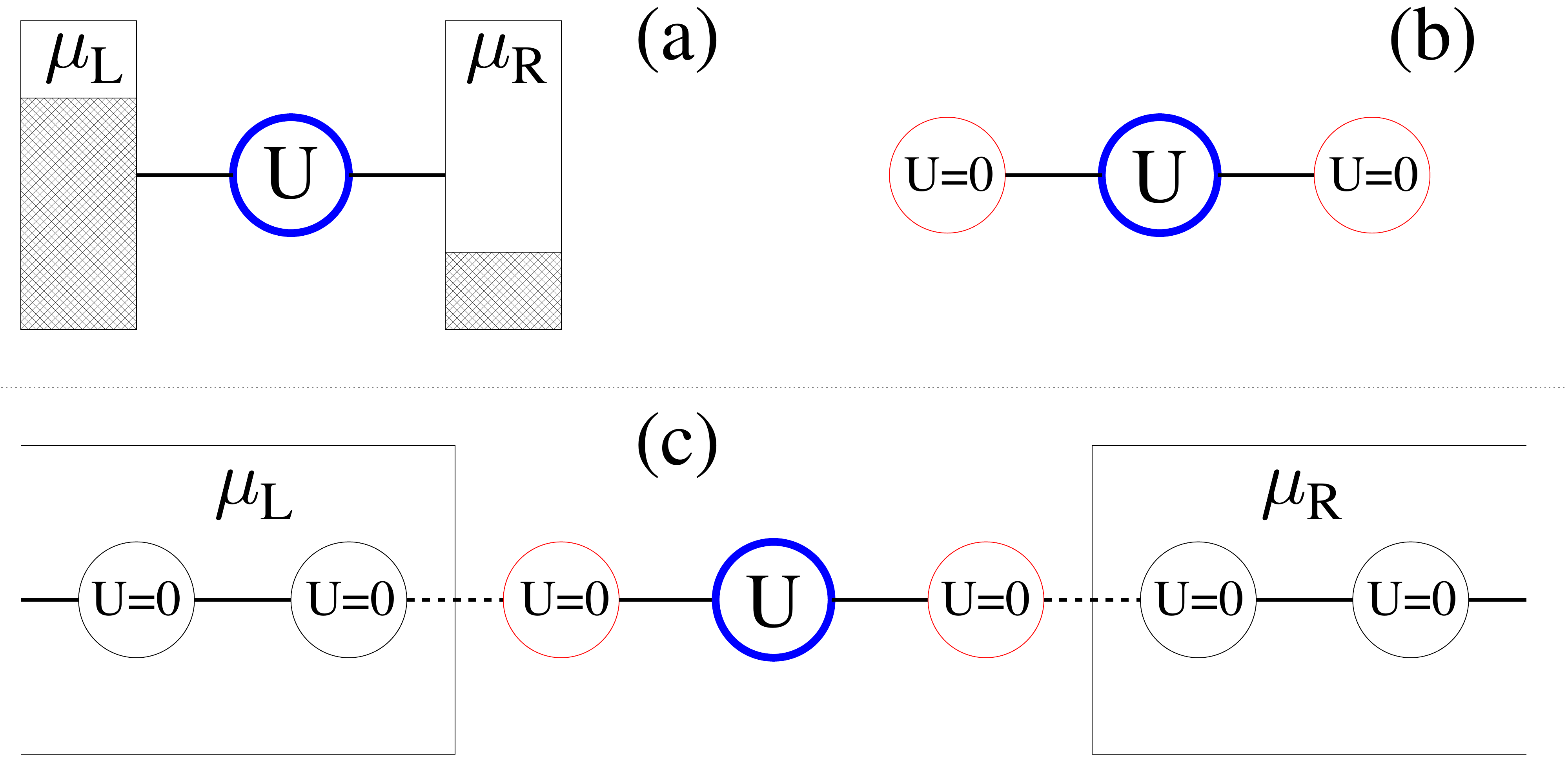}
\caption{\label{fig1}
Nonequilibrium junction model. Shown are
(a) Anderson impurity model;
(b) Reference system within original DF approach~\cite{jung_dual-fermion_2012}; and 
(c) Reference system within auxiliary QME-DF approach. 
}
\end{figure}

One of methods suggested as a way to account for nonlocal correlations
is the dual-fermion (DF) approach~\cite{rubtsov_dual_2008}
(note recent review on diagrammatic methods beyond DMFT~\cite{rohringer_diagrammatic_2018}).
Originally, the method was formulated for equilibrium systems and its efficient implementations 
were reported in the literature~\cite{hafermann_superperturbation_2009,antipov_opendf_2015}.
We note that original DF is a way to account for nonlocal correlations in extended systems.
Later, a nonequilibrium version of the method (DF-inspired superperturbation theory) 
was proposed in Ref.~\onlinecite{jung_dual-fermion_2012} as a way to solve impurity/transport problems.
An attractive feature of the latter formulation is its applicability in the absence of small parameter.
At the heart of consideration is {\em a reference system}, which includes the molecule and
finite number of states representing contacts. Such finite problem can be solved exactly,
however system-baths couplings of the original problem cannot be reproduced properly
due to approximate treatment of the baths. Dual-fermions 
introduce auxiliary zero order Hamiltonian, around which usual
diagrammatic formulation can be formulated. The resulting expansion accounts
for difference between the true system-bath hybridization and its approximation within
the reference system (see Ref.~\onlinecite{jung_dual-fermion_2012} for details).

In situations when one is interested in steady-state behavior, the nonequilibrium DF 
approach of Ref.~\onlinecite{jung_dual-fermion_2012} requires significant numerical effort.
Because only several sites are chosen to represent infinite baths in the reference system,
hybridization function is quite different from the true one. Also, in accordance with the standard
NEGF setup, time propagation starts form decoupled system and baths, so that long time
propagation is required to reach steady-state. Besides, finite size of the reference system
yields necessarily periodic solution, which makes reaching steady-state even harder.

We propose to utilize solution of an auxiliary quantum master equation as a 
reference system for the DF approach in steady-state situations
(compare Figs.~\ref{fig1}b and c).
Auxiliary QME yields description of hybridization and nonequilibrium state of the system 
which are much closer 
to the true solution than any choice of a finite reference system. It also allows to avoid time propagation
in the DF scheme. We note that previously auxiliary QME was proposed in the literature as
a simplified impurity solver for the DMFT calculations~\cite{arrigoni_nonequilibrium_2013,dorda_auxiliary_2014}.
Here we propose to use it as a starting point for a more elaborate (and more accurate)
impurity solver - the nonequilibrium DF approach.


{\bf Nonequilibrium DF.}
For completeness we start with a short description of the nonequilibrium DF approach
(for details see original introduction of the method in Ref.~\onlinecite{jung_dual-fermion_2012}).
One considers reduced dynamics of an open system with interactions confined to
the system subspace and effect of baths degrees of freedom entering via corresponding
self-energies. Effective action of the system defined on the Keldysh contour is
\begin{equation}
\label{S}
S[d^{*},d] = \sum_{1,2} d^{*}_1\,\big[G_0^{-1}-\Sigma^B\big]_{12}\, d_2 + S^{int}[d^{*},d]
\end{equation}
where $i=(m_i,\tau_i)$ ($i=1,2$) is the index incorporating molecular orbital $m_i$ and Keldysh contour
variable $\tau_i$; sum over index implies sum over molecular orbitals and integral over the contour.
$d^{*}_i=d^{*}_{m_i}(\tau_i)$ ($d_i=d_{m_i}(\tau_i)$) is the Grassmann variable corresponding to
creation (annihilation) operator $\hat d_{m_i}^\dagger(\tau_i)$ ($\hat d_{m_i}(\tau_i)$)
of an electron in orbital $m_i$ in the Heisenberg picture~\cite{negele_quantum_1988}.
$G_0^{-1}$ id the inverse free Green function~\cite{wagner_expansions_1991}
\begin{align}
\label{invG0}
&\big[G_0^{-1}\big]_{12}\equiv \delta(\tau_1,\tau_2)\big[i\overset{\rightarrow}{\partial}{}_{\tau_1}\delta_{m_1,m_2} 
- H^0_{m_1m_2}(\tau_1)\big] -
\Sigma^{irr}_{12}
\\
 &\qquad\quad\ \ =\big[-i\overset{\leftarrow}{\partial}{}_{\tau_2}\delta_{m_1,m_2}
 -H^0_{m_1m_2}(\tau_2)\big]\delta(\tau_1,\tau_2)
-\Sigma^{irr}_{12},
\nonumber
\end{align}
 and $\Sigma^B(\tau_1,\tau_2)$ is the self-energy due to coupling to contacts
 \begin{equation}
 \label{SigmaB}
 \Sigma^B_{m_1m_2}(\tau_1,\tau_2) = \sum_{k\in B} V_{m_1k} g_k(\tau_1,\tau_2) V_{km_2}
 \end{equation} 
In Eqs.~(\ref{invG0}) and (\ref{SigmaB}),  $H^0_{m_1m_2}(\tau)$ is the non-interacting part of 
the molecular Hamiltonian, $\Sigma^{irr}_{m_1m_2}(\tau_1,\tau_2)\sim\delta(\tau_1,\tau_2)$ is 
the irregular self-energy, $V_{mk}$ is the matrix element for electron transfer between molecular orbital 
$m$ and contact state $k$, and  
$g_k(\tau_1,\tau_2)\equiv -i \langle T_c\,\hat c_k(\tau_1)\,\hat c_k^\dagger(\tau_2)\rangle$
 is the Green function of free electron in state $k$ of the contacts.
 All intra-molecular interactions are within (unspecified) contribution to the action, $S^{int}[d^{*},d]$.
 
DF approach is based on two important steps. First, one introduces an exactly solvable 
reference system with baths represented by finite number of states. Its known action $\tilde S[d^{*},d]$ 
has the same general form (\ref{S}) with true self-energy $\Sigma^B$ substituted by its
approximate representation $\tilde \Sigma^B$.
So that the desired action $S$ can be written as
\begin{equation}
\label{Srs}
S[d^{*},d]=\tilde S[d^{*},d]+\sum_{1,2} d_1^{*}\,\big[\tilde\Sigma^B-\Sigma^B\big]_{12}\, d_2
\end{equation}
Second, direct application of standard diagrammatic expansion around the interacting reference system
is not possible, because the Wick's theorem is not applicable~\cite{fetter_quantum_1971}. 
To resolve the issue an artificial particle ({\em dual fermion})
is introduced which is used to unravel the term via the Hubbard-Stratonovich transformation~\cite{coleman_introduction_2015}.
Integrating out molecular fermions ($d$ and $d^{*}$) and comparing 
second order cumulant expansion of the resulting expression 
with general form of action for dual fermions,
$S^{DF}[f^{*},f] = \sum_{1,2}f_1^{*}\, \big[\big(G_0^{DF})^{-1}-\Sigma^{DF}\big]_{12}\, f_2$, one gets
\begin{align}
\label{G0DF}
\big(G^{DF}_0\big)^{-1}_{12} &= -g^{-1}_{12} -\sum_{3,4} g^{-1}_{13}\,\big[\tilde\Sigma^B-\Sigma^B\big]^{-1}_{34}\, g^{-1}_{42}
\\
\label{SigmaDF}
\Sigma^{DF}_{12} &= \sum_{3,4} \Gamma_{13;24}\, \big[G_0^{DF}\big]_{43}
\end{align}
Here $g_{12}$ and $\Gamma_{13;24}$ are the single-particle Green function and the two-particle vertex 
of the reference system, respectively~\cite{stefanucci_nonequilibrium_2013}.

Once $\big(G^{DF}\big)= \big[\big(G^{DF}_0\big)^{-1} -\Sigma^{DF}\big]^{-1}$ is known,
single-particle Green function of the molecule is obtained from
\begin{equation}
 \label{G}
 G=\big(\delta\Sigma^B\big)^{-1} + \big[g\,\delta\Sigma^B\big]^{-1}\, G^{DF} \big[\delta\Sigma^B\, g\big]^{-1} 
\end{equation}
where $\delta\Sigma^B\equiv\tilde\Sigma^B-\Sigma^B$.

{\bf Auxiliary QME.}
Choice of the reference system is arbitrary, and the better is ability of the reference system to describe
the original nonequilibrium problem, the better will be also result by the DF approach. 
In this sense choice of a finite number of states to represent baths (see Fig.~\ref{fig1}b) may not be optimal:
its inability to represent properly bath induced dissipation and inevitably periodic solution
makes reaching a steady-state problematic.  We suggest to describe baths with infinite number of states
with majority treated implicitly (integrated out) and including finite number into an extended system 
(see Fig.~\ref{fig1}c). Effectively, this complements choice of Ref.~\onlinecite{jung_dual-fermion_2012}
with actual baths. We suggest using a Markov QME,
\begin{equation}
\label{qme}
 \frac{d\rho^S(t)}{dt} = -i\mathcal{L}\rho^S(t),
\end{equation}
as a tool to simulate the extended system.
Here $\rho^S(t)$ is the extended system density operator and $\mathcal{L}$ is the Liouvillian.
Thus, our analysis will keep all the advantages of the original choice (Hamiltonian of the extended system
is diagonalized exactly), while adding infinite baths, which will allow much more accurate description
of nonequilibrium state and hybridization function of the original system.
Below we focus on steady-state situation, where all correlation functions depend on time difference
and Fourier transform yields their convenient representation in the energy space. 

The nonequilibrium DF approach, Eqs.~(\ref{G0DF})-(\ref{SigmaDF}), requires single- and two-particle Green 
functions of the reference system as an input. To provide those we utilize the quantum regression
relation~\cite{breuer_theory_2003}
\begin{align}
\label{qrr}
& \big\langle T_c\, \hat A(\tau_1)\,\hat B(\tau_2)\ldots \hat Z(\tau_n)\big\rangle =
\\ &\qquad\quad
 \mbox{Tr}\big[\mathcal{O}_n\,\mathcal{U}(t_n,t_{n-1})\ldots
 \mathcal{O}_2\,\mathcal{U}(t_2,t_1)\,\mathcal{O}_1\,\mathcal{U}(t_1,0)\,\rho^S(0)\big]
 \nonumber
\end{align}
Here $\rho^S(0)$ is the steady-state density matrix of the extended system,
$\mathcal{U}(t_i,t_{i-1})$ is the Liouville space evolution operator and
times $t_i$ are ordered so that $t_n>t_{n-1}>\ldots>t_2>t_1>0$.
$\mathcal{O}_i$ is the Liouville space super-operator corresponding to 
one of operators  $\hat A\ldots \hat Z$ whose time is $i$-th in the ordering.
It acts form the left (right) for the operator on the forward (backward) branch of the contour
Steady-state density matrix is found as right eigenvector $\lvert R_0\gg$ corresponding to the Liouvillian eigenvalue $\lambda_0=0$.
Using spectral decomposition of the Liouvillian, evolution operator can be presented in its eigenbasis as
\begin{equation}
\label{U}
 \mathcal{U}(t_i,t_{i-1}) = \sum_\gamma \lvert R_\gamma\gg\, e^{-i\lambda_\gamma(t_i-t_{i-1})}\,
 \ll L_\gamma\rvert
\end{equation}
Note that for evaluation of single- and two-particle Green functions 
besides $\mathcal{L}$ of Eq.~(\ref{qme})  we will have to consider also 
Liouvillians $\mathcal{L}^{(\pm 1)}$ and $\mathcal{L}^{(\pm 2)}$.
These are evolution operator generators for Liouville space vectors $\lvert S_1S_2\gg$ with different 
number $N_S$ of electrons in states $\lvert S_1\rangle$ and $\lvert S_2\rangle$.
For example, for $\mathcal{L}^{(+1)}$, $N_{S_1}=N_{S_2}+1$.
Note, constructing the Liouvillians is helped by conservation of $N_{S_1}-N_{S_2}$ during
evolution. Other symmetries (charge, spin) may help in understanding block structure
within the Liouvillians~\cite{leijnse_kinetic_2008}.

Using (\ref{U}) in (\ref{qrr}) yields expressions for the Green functions of the reference system.
Details of evaluation for single- and two-particle Green functions are given in 
the Appendix.
Once single- and two-particle Green functions of the reference system are known, 
the vertex required in (\ref{SigmaDF}) is given by
\begin{align}
  &\Gamma_{13;24} =
  \\ &\quad
  \sum_{\begin{subarray}{c}1',2'\\3',4'\end{subarray}} 
  g^{-1}_{11'}\, g^{-1}_{33'}
  \big[g^{(2)}_{1'3';2'4'}-g_{1'2'}\,g_{3'4'}+g_{1'4'}\, g_{3'2'}\big]\, g^{-1}_{2'2}\, g^{-1}_{4'4}
  \nonumber
\end{align}
Below we consider extended system of size small enough so that exact diagonalization can be employed.
For system of bigger size more advanced methods (e.g., considerations utilizing matrix product 
states~\cite{dorda_auxiliary_2015}) should be used. 

{\bf Model.}
We apply the QME-DF method to the Anderson impurity model: junction is constructed from quantum dot
coupled to two paramagnetic leads each at its own equilibrium (see Fig.~\ref{fig1}a). 
Hamiltonian of the model is
\begin{equation}
 \hat H = \hat H_M + \sum_{K=L,R}\big( \hat H_K+\hat V_{MK} \big)
\end{equation}
where 
$\hat H_M=\sum_{\sigma=\uparrow,\downarrow} \epsilon_0\, \hat d_\sigma^\dagger\hat d_\sigma
+ U \hat n_{\uparrow}\hat n_{\downarrow}$ and 
$\hat H_K=\sum_{k\in K}\sum_{\sigma=\uparrow,\downarrow}\epsilon_k\, \hat c_{k\sigma}^\dagger\hat c_{k\sigma}$ 
are Hamiltonians of the quantum dot and contact $K$;
$\hat V_{MK}=\sum_{k\in K}\sum_{\sigma=\uparrow,\downarrow}\big( V_{k} \hat d_\sigma^\dagger\hat c_{k\sigma} + H.c.\big)$ describes electron transfer between the dot and contact. 
Here $\hat d_\sigma^\dagger$ ($\hat d_\sigma$) and $\hat c_{k\sigma}^\dagger$ ($\hat c_{k\sigma}$)
creates (annihilates) electron of spin $\sigma$ on the dot and in state $k$ of the contacts, respectively.
$U$ is the Coulomb repulsion and $\hat n_\sigma=\hat d_\sigma^\dagger\hat d_\sigma$.

We use the QME-DF, Eq.~(\ref{G}), to calculate single-particle Green function of the dot 
\begin{equation}
G_\sigma(\tau_1,\tau_2) = -i\langle T_c\,\hat d_\sigma(\tau_1)\,\hat d_\sigma^\dagger(\tau_2)\rangle
\end{equation}
and use it in steady-state simulations of level population $n_\sigma$,  spectral function $A_\sigma(E)$,
and current $I_L=-I_R$~\cite{jauho_time-dependent_1994} 
\begin{equation}
\begin{split}
 & n_\sigma = -i \int\frac{dE}{2\pi}G^{<}_\sigma(E);
 \quad A_\sigma(E) = -\frac{1}{\pi}\mbox{Im}\, G^r_\sigma(E);
  \\
 & I_K = \sum_\sigma \int\frac{dE}{2\pi} \big(\Sigma^{<}_K(E)\,G_\sigma^{>}(E)-\Sigma^{>}_K(E)\, G^{<}_\sigma(E)\big)
 \end{split}
 \end{equation}
Here $<$, $>$ and $r$ are lesser, greater and retarded projections, respectively.
$\Sigma^{\gtrless}_K(E)$ is greater/lesser projection of self-energy due to coupling to contact $K$
($L$ or $R$).

We model contacts as semi-infinite tight-binding chains with on-site energies $\epsilon_K$
and hopping parameter $t_K$ ($K=L,R$); electron hopping between quantum dot and the chain
is $t_{MK}$. This is the Newns-Anderson model~\cite{newns_self-consistent_1969}.


\begin{figure}[t]
\centering\includegraphics[width=\linewidth]{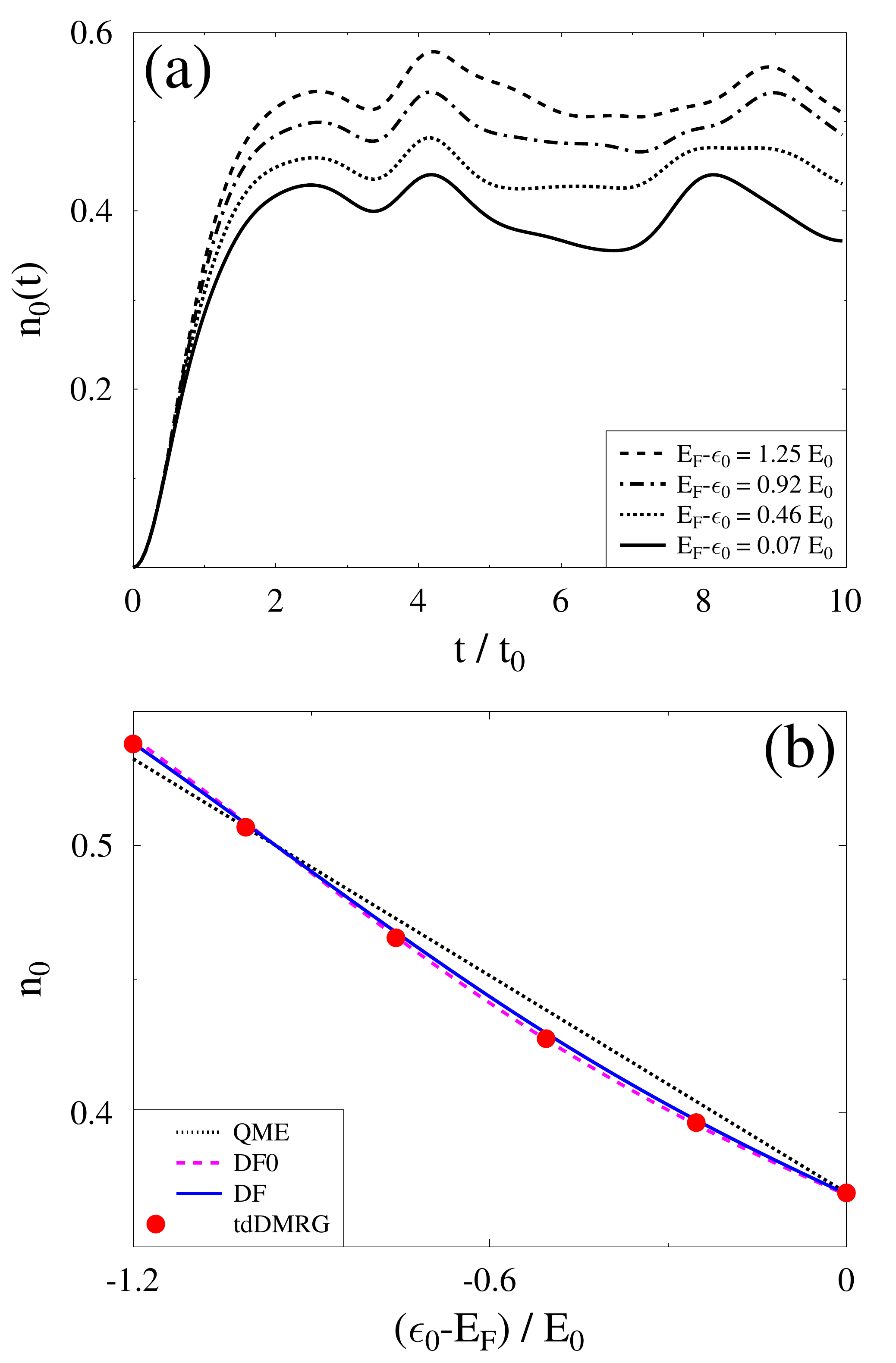}
\caption{\label{fig2}
(Color online)
Population of molecular level in biased junction.
Shown are
(a) Results of original nonequilibrium DF simulation,
where at $t=0$ coupling between system and contacts is switched on,
for a set of level positions;
(b) Dependence of steady-state level population on position of the level 
calculated from auxiliary QME (dotted line, black), zero (dashed line, magenta) and first (solid line, blue)
QME-DF approaches. Circles (red) represent results of numerically exact tdDMRG simulations. See text for parameters.
}
\end{figure}

{\bf Numerical results.}
Here we apply the QME-DF approach to the Anderson impurity model and compare its results with
the original nonequilibrium DF scheme and with numerically exact tdDMRG calculations.
The latter was performed by ALPS-MPS~\cite{bauer_alps_2011,dolfi_matrix_2014}.  
We show two flavors of the QME-DF results: zero order, when one neglects self-energy $\Sigma^{DF}$,
and first order when the self-energy is evaluated according to Eq.~(\ref{SigmaDF}). 
Parameters and results of of the simulations are presented in terms of maximum total escape rate,
$\Gamma_0=2\, t_{ML}^2/t_L+2\, t_{MR}^2/t_R$, and units derived from it. 
In particular, we employ units of energy, $E_0=\Gamma_0$, time $t_0=\hbar/E_0$, bias $V_0=E_0/|e|$,
and current, $I_0=|e| E_0/\hbar$. 

Unless stated otherwise parameters of the simulations are as follows:
$U=5\, E_0$, $\epsilon_0=-U/2$, $t_{ML}=t_{MR}=0.79 E_0$ and $t_L=t_R=2.5 E_0$.
Positions of on-site energies in contacts, $\epsilon_L$ and $\epsilon_R$, are defined by corresponding 
chemical potentials, $\mu_L$ and $\mu_R$.  Fermi energy is taken as origin, $E_F=0$,
and bias is assume dot be applied symmetrically, $\mu_L=E_F+|e|V_{sd}/2$ and $\mu_R=E_F-|e|V_{sd}/2$.
Simulations are performed at zero temperature on energy grid spanning range from 
$-12.5 E_0$ to $12.5 E_0$ with step $0.0125 E_0$. 
  
\begin{figure}[t]
\centering\includegraphics[width=\linewidth]{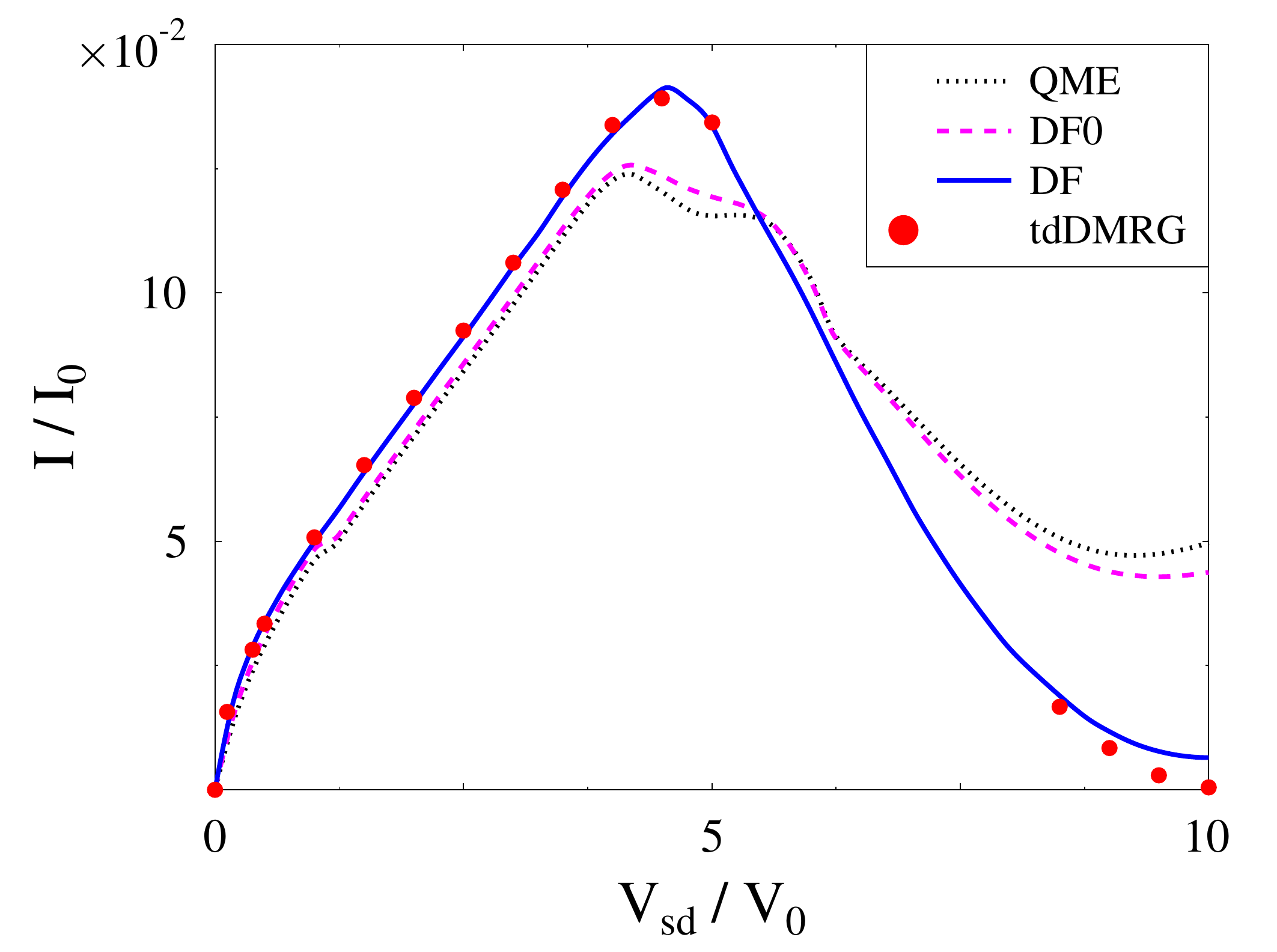}
\caption{\label{fig3}
(Color online) Current voltage characteristics. Shown are results of
the auxiliary QME (dotted line, black), zero (dashe dline, magenta)
and first (solid line, blue) order QME-DF approaches.
Circles (red) represent results of numerically exact tdDMRG simulations.
See text for parameters.
}
\end{figure}
  
Figure~\ref{fig2} shows results of simulation of level population $n_\uparrow=n_\downarrow\equiv n_0$
for junction under bias $V_{sd}=2.5 V_0$. Here $U=2 E_0$ and simulations are done for a set of level positions.
Fig.~\ref{fig2}a demonstrates time propagation of the population
after attaching quantum dot to contacts at $t=0$ simulated within the nonequilibrium DF simulation of 
Ref.~\cite{jung_dual-fermion_2012}. 
Note that time propagation is at the heart of the original approach.
One sees that reaching steady-state is indeed quite problematic within the approach.
Fig.~\ref{fig2}b shows level population calculated within zero (DF0, dashed line) and first (DF, solid line) order  
QME-DF approach vs. level position. For comparison we show auxiliary QME (QME, dotted line) and
numerically exact tdDMRG results. 
One sees, that QME-DF approach is quite accurate in predicting level population.
Note that QME-DF simulation for steady-state is much easier (both numerically and in terms of getting 
the steady state result) than original DF formulation.

\begin{figure}[t]
\centering\includegraphics[width=\linewidth]{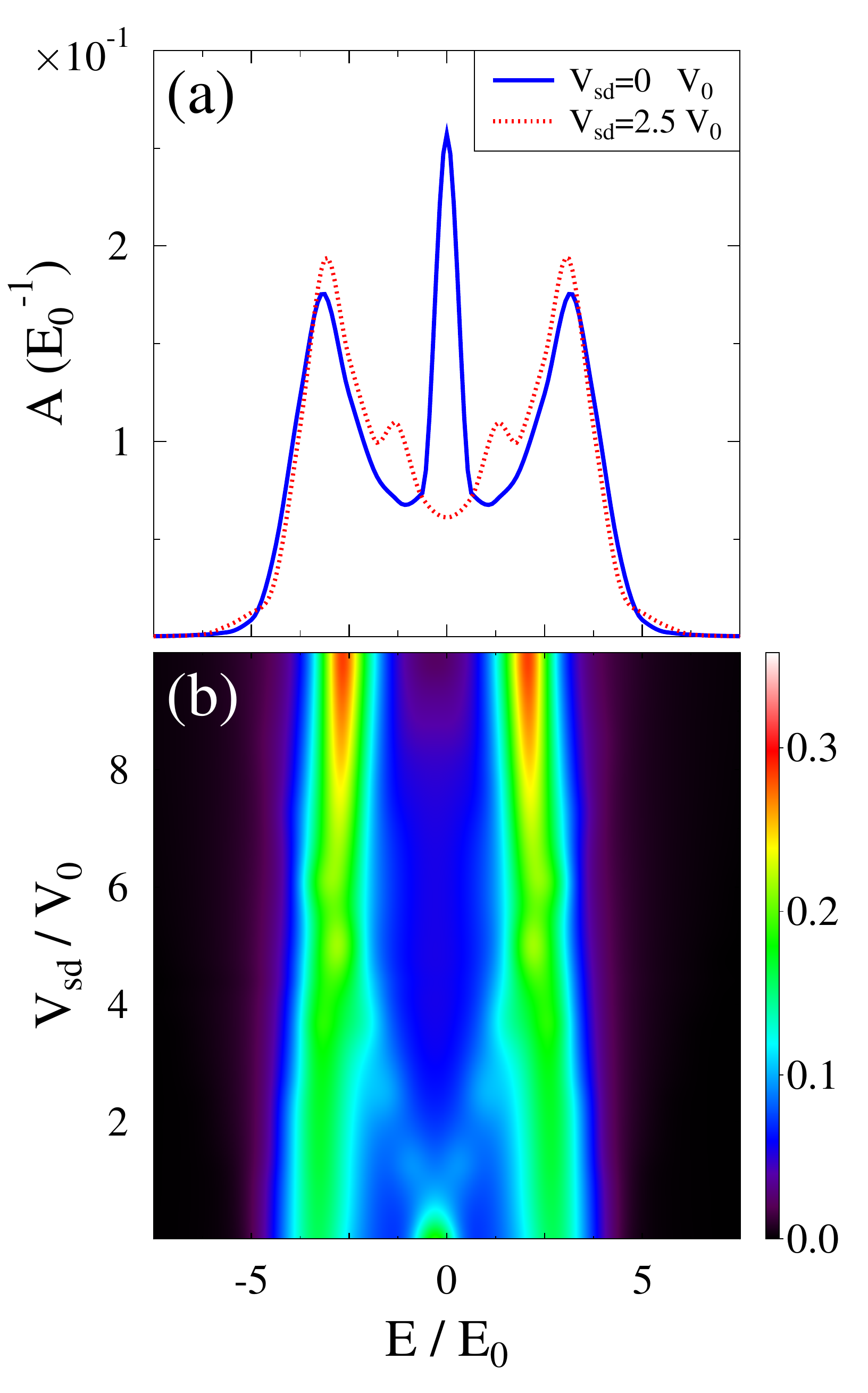}
\caption{\label{fig4}
(Color online)
Spectral function of Anderson impurity model. Shown are 
results of the QME-DF simulations for
(a) Spectral function of unbiased junction (solid line, blue)
and non-equilibrium situation with $V_{ѕd}/V_0=2.5$ (dotted line, red);
and 
(b) Map of spectral function vs. energy and applied bias.
See text for parameters.
}
\end{figure}

Current-voltage characteristics of the Anderson impurity model is shown in Figure~\ref{fig3}.
Here we show results of the auxiliary QME (Dotted line), the zero (dashed line)
and first (solid line) order QME-DF. 
The latter is quite close to the exact tdDMRG results (circles). The results are taken from Ref.~\onlinecite{dorda_auxiliary_2014}

It is interesting to note that first order QME-DF  calculation with three auxiliary sites employed
for the auxiliary QME simulation yields result similar to one reachable for six auxiliary sites in
QME simulation (compare with Fig.~3 of Ref.~\cite{dorda_auxiliary_2014}). 

Finally, we show results of simulations of spectral function. Fig.~\ref{fig4}a shows spectral function
of junction at equilibrium (solid line) and under bias of $2.5 V_0$. Spectral function demonstrates 
a Kondo like feature which is splitted in biased junction. Similar results were obtained in
Refs.~\cite{meir_low-temperature_1993,wingreen_anderson_1994,sivan_single-impurity_1996}.
This behavior is only qualitative representation of a true Kondo behavior.
To improve numerical accuracy one has to account for higher order diagrams as done, e.g.,
in Ref.~\cite{hafermann_superperturbation_2009} in equilibrium DF consideration.
Fig.~\ref{fig4}b shows change of spectral function with bias. At low biases equilibrium Kondo peak splits
and follows corresponding chemical potentials, while higher biases destroy the correlation.


{\bf Conclusion.}
The nonequilibirum dual fermion approach introduced originally in Ref.~\cite{jung_dual-fermion_2012}
is a promising method for simulating strongly correlated open systems.
Contrary to usual diagrammatic expansions in small interaction (intra-system interaction
in, e.g., NEGF or system-bath couplings in, e.g., PP- or Hubbard NEGF)
the method is capable to treat systems in the absence of a small parameter
by considering expansion around an exactly solvable reference system, 
which includes the system of interest and a finite number of states representing contacts. 
Choice of a finite reference system in the original DF formulation results in periodic dynamics,
which together with necessity to consider time propagation starting from decoupled system and baths
and the fact that finite number of states cannot properly describe bath induced dissipation
complicates simulation of steady-states.

We proposed complementing finite reference system with infinite baths and use auxiliary quantum master
equation as a tool for its solution. We argued that the approach is advantageous in treating
steady-states because it yields reference system which is much closer to the true nonequilibirum state
than in the original formulation. Besides, infinite size of the modified reference system
results in more accurate description of bath induced dissipation. Finally, the approach allows to
avoid long time propagations necessary to reach steady-state solution in the original formulation. 
We used the Anderson impurity as a test model and compared the QME-DF simulations with
numerically exact tdDMRG results. The new scheme is shown to be quite accurate
and relatively inexpensive numerically.
Further development of the method and its application to realistic systems is a goal for future research.


\begin{acknowledgments}
M.G. acknowledges support by the National Science Foundation  (grant CHE-1565939).
G.C. acknowledges support by the Israel Science Foundation (Grant No. 1604/16).
\end{acknowledgments}

\appendix
\section*{Appendix: Evaluation of single- and two-particle Green functions of reference system}

To evaluate dual-fermion self-energy, one has to calculate the two-particle vertex, Eq.(6).
The latter depends on single- and two-particle Green functions of the reference system, Eq.(11).
The Green functions are obtained by employing the quantum regression relation. 

\subsection*{Evaluation of single-particle Green functions}
Here we discuss details of evaluation of single-particle Green function 
\begin{equation}
 g_{12} = -i\langle T_c \hat d_1\, \hat d_2^\dagger \rangle_{ref}
\end{equation}
where as previously $\hat d_i=\hat d_{m_i}(\tau_i)$, and 
where subscript `ref' indicates quantum mechanical and statistical average with respect 
to total density matrix of the reference system.
Evaluation of the Green function requires consideration of four projections
\begin{align}
g^{\gtrless\,+}_{m_1m_2}(t_1,t_2) &\equiv \theta(t_1-t_2) g_{m_1m_2}^{\gtrless}(t_1,t_2)
\\
g^{\gtrless\,-}_{m_1m_2}(t_1,t_2) &\equiv \theta(t_2-t_1) g_{m_1m_2}^{\gtrless}(t_1,t_2)
\end{align}
where 
\begin{align}
g_{m_1,m_2}^{>}(t_1,t_2) &=-i\langle \hat d_{m_1}(t_1)\hat d_{m_2}^\dagger(t_2)\rangle_{ref}\\
g_{m_1,m_2}^{>}(t_1,t_2) &=i\langle \hat d_{m_2}^\dagger(t_2)\hat d_{m_1}(t_1)\rangle_{ref}.
\end{align}
Fourier transforms of the four projections are
\begin{widetext}
\begin{align}
g_{m_1m_2}^{>\,+}(E) &=\quad \sum_\gamma\sum_{\{S_i\},\{S_i'\}}
\big[\xi^{m_1}_{S_3S_3'}\big]^{*}\, \xi^{m_2}_{S_2S_1}\,\rho^S_{S_1S_1'}
\frac{\ll S_3S_3'\vert R_\gamma^{(+1)}\gg \, \ll L_\gamma^{(+1)}\vert S_2S_1'\gg}
{E-\lambda_\gamma^{(+1)}}
\\
g_{m_1m_2}^{<\,+}(E) &= -\sum_\gamma\sum_{\{S_i\},\{S_i'\}}
\big[\xi^{m_1}_{S_3S_3'}\big]^{*}\, \xi^{m_2}_{S_1'S_2'}\,\rho^S_{S_1S_1'}
\frac{\ll S_3S_3'\vert R_\gamma^{(+1)}\gg \,\ll L_\gamma^{(+1)}\vert S_1S_2'\gg}
{E-\lambda_\gamma^{(+1)}}
\\
g_{m_1m_2}^{>\,-}(E) &= -\sum_\gamma\sum_{\{S_i\},\{S_i'\}}
\xi^{m_2}_{S_3'S_3}\, \big[\xi^{m_1}_{S_2'S_1'}\big]^{*}\, \rho^S_{S_1S_1'}
\frac{\ll S_3S_3'\lvert R_\gamma^{(-1)}\gg\, \ll L_\gamma^{(-1)} \vert S_1S_2'\gg}
{E+\lambda_\gamma^{(-1)}}
\\
g_{m_1m_2}^{<\,-}(E) &=\quad \sum_\gamma\sum_{\{S_i\},\{S_i'\}}
\xi^{m_2}_{S_3'S_3}\, \big[\xi^{m_1}_{S_1S_2}\big]^{*}\,\rho^S_{S_1S_1'}
\frac{\ll S_3S_3' \vert R_\gamma^{(-1)}\gg \,\ll L_\gamma^{(-1)} \vert S_2S_1' \gg}
{E+\lambda_\gamma^{(-1)}}
\end{align}
\end{widetext}
Here $\xi^m_{S_2S_1}\equiv \langle S_2\rvert \hat d_m^\dagger \lvert S_1\rangle$;
$\lambda_\gamma^{(\pm 1)}$, $\lvert R_\gamma^{(\pm 1)}\gg$ and $\ll L_\gamma^{(\pm 1)}\rvert$
are eigenvalue, right and left eigenvectors of the Liouvillian $\mathcal{L}^{(\pm 1)}$.

\subsection*{Evaluation of two-particle Green functions}
Here we provide details of evaluation for two-particle Green function
\begin{equation}
\label{def_g2}
g^{(2)}_{13,24}=-\langle T_c\, \hat d_1\,\hat d_3\,\hat d_4^\dagger\,\hat d_2^\dagger\rangle
\end{equation}
To connect (\ref{def_g2}) to Liouville QME formulation, one has to consider $2^4=16$ projections of the Green function on the contour 
and $4!=24$ time orderings.  

It is convenient to introduce Liouville space matrix elements of electron annihilation operators at
time-ordered ($s=0$ or $-$) and anti-time ordered ($s=1$ or $+$) branches of the Keldysh contour
\begin{equation}
\label{d_me}
\ll S^2_{-}S^2_{+}\rvert \hat d_m^s \lvert S^1_{-}S^1_{+}\gg =
\begin{cases}
\delta_{S^2_{+},S^1_{+}} \langle S^2_{-} \rvert \hat d_m \lvert S^1_{-}\rangle & s=0
\\
\delta_{S^2_{-},S^1_{-}} \langle S^1_{+} \rvert \hat d_m \lvert S^2_{+}\rangle & s=1
\end{cases}
\end{equation}
Similar  definitions hold for creation operators.

Explicit form of quantum regression relation for two particle Green function will depend on time ordering in (\ref{def_g2}).
For example, for $t_4>t_3>t_2>t_1$ and indicating contour projections of operators by respectively $s_4$, $s_3$, $s_2$, and $s_1$
expression for two-particle Green function in terms of QME solution (Liouvillian eigenvalues $\lambda_\gamma$ and
eigenvectors $\lvert R_\gamma\gg$ and $\ll L_\gamma\rvert$) is
\begin{widetext}
\begin{align}
&-\sum_{\{\gamma_i\}} (-1)^p e^{-i\lambda^{(-1)}_{\gamma_3}(t_4-t_3)} \,
e^{-i\lambda^{(0)}_{\gamma_2}(t_3-t_2)} \, e^{-i\lambda^{(-1)}_{\gamma_3}(t_2-t_1)}
\times \sum_{S}\prod_{\{s_i=\mp\}}\sum_{\{S^j_{s_i}\}}
\nonumber \\ &
\label{SM_example}
\ll S\, S\rvert d^{\dagger\, s_4}_{m_4} \lvert S^7_{-}S^7_{+}\gg\, \ll S^7_{-}S^7_{+}\vert R_{\gamma_3}^{(-1)}\gg\,
\ll L_{\gamma_3}^{(-1)}\vert S^6_{-}S^6_{+}\gg\, \ll S^6_{-}S^6_{+}\rvert d^{s_3}_{m_3}\lvert S^5_{-}S^5_{+}\gg\,
\ll S^5_{-}S^5_{+} \vert R^{(0)}_{\gamma_2} \gg\, 
\\ &
\ll L^{(0)}_{\gamma_2} \vert S^4_{-} S^4_{+}\gg\,
\ll S^4_{-} S^4_{+} \rvert d_{m_2}^{\dagger\, s_2} \lvert S^3_{-}S^3_{+} \gg\,
\ll S^3_{-}S^3_{+} \vert R_{\gamma_1}^{(-1)} \gg\, \ll L_{\gamma_1}^{(-1)} \vert S^2_{-}S^2_{+}\gg\,
\ll S^2_{-}S^2_{+} \rvert d_{m_1}^{s_1} \lvert S^1_{-}S^1_{+}\gg\, \rho^{S}_{S^1_{-}S^1_{+}}
\nonumber 
\end{align}
where $p$ is permutation of creation and annihilation operators ($\hat d$ and $\hat d^\dagger$) in the projection.
Similar expressions can be written to other $23$ time orderings.
\end{widetext}

Because time dependence is explicit in (\ref{SM_example}), time integrals in Eq.(6) can be evaluated 
analytically. Also, terms in second and third rows of (\ref{SM_example}) can be combined into groups in which 
matrix products may be pre-calculated only once and stored in memory. 
Specifically for (\ref{SM_example}) second and third rows of the expression can be 
presented as product of two matrices and two vectors
\begin{equation}
\label{SM_expr}
 v_{\gamma_3}\, M_{\gamma_3\gamma_2}\, N_{\gamma_2\gamma_1}\, w_{\gamma_1}
\end{equation}
where
\begin{widetext}
\begin{align}
\label{SM_vecmat}
v_{\gamma_3} &= \sum_{S,S^7_{-},S^7_{+}}
\ll S\, S\rvert d^{\dagger\, s_4}_{m_4} \lvert S^7_{-}S^7_{+}\gg\, \ll S^7_{-}S^7_{+}\vert R_{\gamma_3}^{(-1)}\gg
\\
M_{\gamma_3\gamma_2} &= \sum_{S^6_{-},S^6_{+},S^5_{-},S^5_{+}} 
\ll L_{\gamma_3}^{(-1)}\vert S^6_{-}S^6_{+}\gg\, \ll S^6_{-}S^6_{+}\rvert d^{s_3}_{m_3}\lvert S^5_{-}S^5_{+}\gg\,
\ll S^5_{-}S^5_{+} \vert R^{(0)}_{\gamma_2} \gg
\nonumber \\
N_{\gamma_2\gamma_1} &= \sum_{S^4_{-},S^4_{+},S^3_{-},S^3_{+}}
\ll L^{(0)}_{\gamma_2} \vert S^4_{-} S^4_{+}\gg\,
\ll S^4_{-} S^4_{+} \rvert d_{m_2}^{\dagger\, s_2} \lvert S^3_{-}S^3_{+} \gg\,
\ll S^3_{-}S^3_{+} \vert R_{\gamma_1}^{(-1)} \gg
\nonumber \\
w_{\gamma_1} &= \sum_{S^2_{-},S^2_{+},S^1_{-},S^1_{+}}
\ll L_{\gamma_1}^{(-1)} \vert S^2_{-}S^2_{+}\gg\,
\ll S^2_{-}S^2_{+} \rvert d_{m_1}^{s_1} \lvert S^1_{-}S^1_{+}\gg\, \rho^{S}_{S^1_{-}S^1_{+}}
\nonumber
\end{align}
\end{widetext}



\end{document}